\documentclass[nofootinbib,twocolumn,prl,preprintnumbers]{revtex4-1}
\pdfoutput=1
\usepackage{amsmath,amsthm,amssymb,multirow,psfrag}
\usepackage{epsfig}
\usepackage{color}

\graphicspath{{./Figures/}}

\begin{document}

%%%%%%%%%%%new definitions: 
\def\lsim{\mathrel{\rlap{\lower4pt\hbox{\hskip1pt$\sim$}}
  \raise1pt\hbox{$<$}}}
\def\gsim{\mathrel{\rlap{\lower4pt\hbox{\hskip1pt$\sim$}}
  \raise1pt\hbox{$>$}}}
\newcommand{\vev}[1]{ \left\langle {#1} \right\rangle }
\newcommand{\bra}[1]{ \langle {#1} | }
\newcommand{\ket}[1]{ | {#1} \rangle }
\newcommand{\ev}{ {\rm eV} }
\newcommand{\kev}{{\rm keV}}
\newcommand{\mev}{{\rm MeV}}
\newcommand{\tev}{{\rm TeV}}
\newcommand{\mpl}{$M_{Pl}$}
\newcommand{\mw}{$M_{W}$}
\newcommand{\Ft}{F_{T}}
\newcommand{\Zparity}{\mathbb{Z}_2}
\newcommand{\BLambda}{\boldsymbol{\lambda}}
\newcommand{\met}{\;\not\!\!\!{E}_T}
\newcommand{\beq}{\begin{equation}}
\newcommand{\eeq}{\end{equation}}
\newcommand{\bea}{\begin{eqnarray}}
\newcommand{\eea}{\end{eqnarray}}
\newcommand{\nn}{\nonumber}
\newcommand{\gev}{{\mathrm GeV}}
\newcommand{\hc}{\mathrm{h.c.}}
\newcommand{\eps}{\epsilon}
\newcommand{\bwt}{\begin{widetext}}
\newcommand{\ewt}{\end{widetext}}
\newcommand{\draftnote}[1]{{\bf\color{blue} #1}}

\newcommand{\cO}{{\cal O}}
\newcommand{\cL}{{\cal L}}
\newcommand{\cM}{{\cal M}}

%References  
\newcommand{\fref}[1]{Fig.~\ref{fig:#1}} 
\newcommand{\eref}[1]{Eq.~\eqref{eq:#1}} 
\newcommand{\aref}[1]{Appendix~\ref{app:#1}}
\newcommand{\sref}[1]{Section~\ref{sec:#1}}
\newcommand{\tref}[1]{Table~\ref{tab:#1}}

\title{\Large{{\bf Golden Probe of Electroweak Symmetry Breaking}}}

\author{{\bf {Yi Chen$\,^{a,b}$,~Joe Lykken$\,^{c}$,~Maria Spiropulu$\,^{a}$,~Daniel Stolarski$\,^{d}$,~Roberto Vega-Morales$\,^{e}$ }}}

\affiliation{
$^a$Lauritsen Laboratory for High Energy Physics, California Institute of Technology, Pasadena,~USA\\
$^b$CERN, European Organization for Nuclear Research, Geneva, Switzerland\\
$^c$Theoretical Physics Department, Fermilab, P.O.~Box 500, Batavia, IL 60510, USA\\
$^d$Ottawa-Carleton Institute for Physics, Carleton University, 1125 Colonel By Drive, Ottawa, Ontario K1S 5B6, Canada\\
$^e$Departamento de F\'{i}sica Te\'{o}rica y del Cosmos and CAFPE,
Universidad de Granada,\\ Campus de Fuentenueva, E-18071 Granada, Spain
}
\preprint{UG-FT 322/16}
\preprint{CAFPE 192/16}
\preprint{FERMILAB-PUB-16-304-PPD}
\email{
yichen@caltech.edu\\
lykken@fnal.gov\\
stolar@physics.carleton.ca\\
smaria@caltech.edu\\
rvegamorales@ugr.es\\}
\begin{abstract}
The ratio of the Higgs couplings to $WW$ and $ZZ$ pairs, $\lambda_{WZ}$, is a fundamental parameter in electroweak symmetry breaking as well as a measure of the (approximate) custodial symmetry possessed by the gauge boson mass matrix.~We show that Higgs decays to four leptons are sensitive, via tree level/1-loop interference effects, to both the magnitude and, in particular, overall sign of $\lambda_{WZ}$.~Determining this sign requires interference effects, as it is nearly impossible to measure with rate information.~Furthermore, simply determining the sign effectively establishes the custodial representation of the Higgs boson.~We find that $h\to4\ell$ ($4\ell \equiv 2e2\mu, 4e, 4\mu$) decays have excellent prospects of directly establishing the overall sign at a high luminosity 13~TeV LHC.~We also examine the ultimate LHC sensitivity in $h\to4\ell$ to the magnitude of $\lambda_{WZ}$.~Our results are independent of other measurements of the Higgs boson couplings and, in particular, largely free of assumptions about the top quark Yukawa couplings which also enter at 1-loop.~This makes $h\to4\ell$ a unique and independent probe of the electroweak symmetry breaking mechanism and custodial symmetry.
\end{abstract}
\maketitle

%%%%%%%%%%%%%%%%%%%%%%%%%%%%%%%%%%
\section{Introduction} \label{sec:intro} 

The precise nature of electroweak symmetry breaking (EWSB) is one of the fundamental questions the LHC is poised to answer.~Measurements of the recently discovered~\cite{:2012gk,:2012gu} Higgs-like boson at 125 GeV are a critical component of this program, and the details of its interactions with the heavy gauge bosons, the $W$ and the $Z$, are of particular importance.~Measurements of Higgs decays indicate that its couplings to $WW$ and $ZZ$ pairs are mediated at tree level via the operators,
\beq
\mathcal{L}_{ZW} \supset
\frac{h}{v} 
\Big( g_{Z} m_Z^2 Z^\mu Z_\mu + 2 g_{W} m_W^2 W^{\mu+} W^-_{\mu} \Big) .
\label{eq:hLag}
\eeq
The couplings $g_{Z}$ and $g_{W}$ depend sensitively on the nature of EWSB.~Though their values are predicted in the Standard Model (SM), other theories of EWSB give different predictions.~The ratio of these couplings, 
\bea
\label{eq:LWZ}
\lambda_{WZ} = g_{W}/g_{Z} ,
\eea
is a particularly useful quantity and gives crucial information on the nature of EWSB and the electroweak properties of the $h(125)$ resonance.

While decay rates can be used to probe the magnitude of $\lambda_{WZ}$, interference effects are needed to probe the overall sign.~Thus measurements by ATLAS and CMS of $h \to WW$ and $h \to ZZ$ decay rates for the 125~GeV resonance~\cite{Aad:2013xqa,Chatrchyan:2013mxa} are sensitive to $|\lambda_{WZ}|$, but they can \textit{not} determine the overall sign.~Combined measurements of the production channels $gg\to tH$ and $gg\to ZH$ allow one to potentially gain sensitivity to the overall sign of $\lambda_{WZ}$~\cite{AtlasTalk} through interference effects, but this requires a fit of the observables in multiple channels as well as various assumptions about the presence (or lack) of effects from physics beyond the SM.~Finally, the rate of these processes is quite small.~It would thus be valuable to have an independent probe free of such considerations and that is directly sensitive to the overall sign of $\lambda_{WZ}$.

We propose to exploit tree level/1-loop interference effects in Higgs boson decays to four charged leptons (electrons and muons) as a new avenue for studying $\lambda_{WZ}$.~The sensitivity to $\lambda_{WZ}$ arises from interference between the tree level $ZZ$ mediated amplitude and one-loop contributions involving the $W$ boson.~Since the interference depends linearly on $\lambda_{WZ}$, this makes $h\to4\ell$ ($4\ell \equiv 2e2\mu, 4e, 4\mu$) decays sensitive to both the magnitude \emph{and} overall sign of $\lambda_{WZ}$.
 
We can also consider the assumption that the Higgs scalar potential respects the well known `custodial' global $SU(2)$ symmetry~\cite{Sikivie:1980hm} as implied by electroweak precision data~\cite{Agashe:2014kda}.~In this case~$\lambda_{WZ}$ depends only on the custodial representation~\cite{Low:2010jp} of $h$, which is restricted to be either a singlet as in the SM, or fiveplet as can be found in the well known Georgi-Machacek (GM) model~\cite{Georgi:1985nv,Chanowitz:1985ug,Gunion:1989ci,Gunion:1990dt} and its supersymmetric incarnation~\cite{Cort:2013foa,Garcia-Pepin:2014yfa,Delgado:2015bwa}.~As shown in~\cite{Low:2010jp}, this allows only two possibilities for the ratio:
\bea\label{eq:lam}
\lambda_{WZ} &=& +1~~\rm{(singlet)} \nn\\
\lambda_{WZ} &=& -1/2~~\rm{(fiveplet)} .
\eea
While there are necessarily custodial symmetry breaking corrections~\cite{Gunion:1989ci,Gunion:1990dt,Garcia-Pepin:2014yfa} to $\lambda_{WZ}$ induced via hypercharge interactions at one loop, as well as Yukawa interactions at one or two loops depending on the electroweak representation of $h$, they are much too small to give the $\mathcal{O}(1)$ corrections needed to change the sign of the tree level predictions in~\eref{lam}.~Therefore, a determination of the sign of $\lambda_{WZ}$ effectively establishes the custodial representation\footnote{Current data favors $|\lambda_{WZ}| = 1$, but the central value is below one and $|\lambda_{WZ}| = 1/2$ is still consistent at the $\sim\,3\,\sigma$ level~\cite{Corbett:2015ksa,Khachatryan:2016vau}.} of the Higgs boson.

In this study, we utilize the $h\to4\ell$ framework developed in~\cite{Gainer:2011xz,Stolarski:2012ps,Chen:2012jy,Chen:2013ejz,Chen:2014gka,Chen:2015iha,Chen:2015rha} which uses all kinematic observables in the \emph{normalized} fully differential decay width to perform a matrix element analysis of the sensitivity to $\lambda_{WZ}$.~We first briefly review how top and $W$ boson loops enter into $h\to4\ell$.~We then perform a likelihood analysis to estimate how much data will be needed to establish the overall sign of $\lambda_{WZ}$ independently of the top Yukawa sector.~Since we are only interested in a ratio of couplings, we can take advantage of this to use only \emph{shape} information.~Thus, while rate information is not used, our analysis is largely independent of the uncertainties and assumptions associated with Higgs production effects.

Under minimal assumptions about the top Yukawa sector, we estimate that a 13~TeV LHC will begin to become sensitive to the overall sign of $\lambda_{WZ}$ in $h\to4\ell$ decays with $\mathcal{O}(800)~\rm{fb}^{-1}$ of data.~We also perform hypothesis testing to estimate how much data will be needed to distinguish between the two predictions of custodial symmetry given in~\eref{lam} as well as estimate how much data will be needed to begin making precision measurements of $\lambda_{WZ}$ at the LHC or future colliders.~In general we find that $h\to4\ell$ decays should serve as a unique and complementary, but independent, probe of EWSB and custodial symmetry at the LHC and beyond.

%%%%%%%%%%%%%%%%%%%%%%%%%%%%%
%%%%%%%%%%%%%%%%%%%%%%%%%%%%%
%%%%%%%%%%%%%%%%%%%%%%%%%%%%%

%%%%%%%%%%%%%%%%%%%%%%%%%%%%%
\section{Loops, Effective Couplings,\\ and Backgrounds}
\label{sec:loops}

Here we briefly review the most relevant aspects of how $\lambda_{WZ}$ enters into the $h\to4\ell$ amplitude through the $hVV$ effective couplings which mediate $h\to4\ell$ decays (see~\fref{hto4l}) at both tree level and one-loop.~A more in depth discussion of these and other one-loop effects contributing to $h\to4\ell$ can be found in~\cite{Chen:2015rha} to which we refer the reader for details.~We also briefly summarize how the top Yukawa sector and dominant $4\ell$ background are incorporated into our analysis.

%%%%%%%%%%%%%%%
\subsection{Loops and effective couplings}

The decay of the Higgs to four leptons is mediated at tree-level by $Z$ boson pairs (see left~\fref{Feyns}), while at one-loop the leading effects are  mediated by $W$ boson loops as shown in~\fref{Feyns} (center).~There is typically a sizable contribution from top quark loops such as in the SM also shown in~\fref{Feyns} (right).~This has been utilized to study the CP properties of the top quark Yukawa sector~\cite{Chen:2015rha} by exploiting one loop/tree level interference effects.~In a similar manner, in this study we utilize one loop/tree level interference effects to probe $\lambda_{WZ}$.
\begin{figure}[tb]
\centering
\begin{minipage}[c]{0.3\textwidth}
\includegraphics[width=1.1\textwidth]{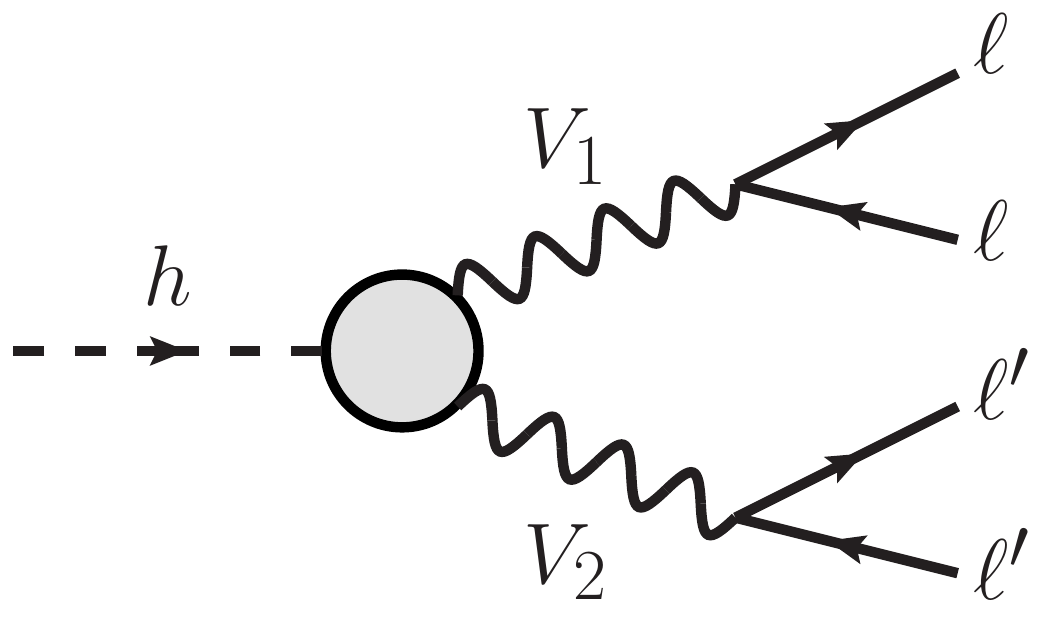}
\end{minipage}
\hfill
\caption{Schematic representation of the $hVV$ contributions to the $h\to4\ell$ amplitude where $V_{1,2} = Z,\gamma$ and $\ell, \ell' = e, \mu$.}
\label{fig:hto4l}
\end{figure}
\begin{figure*}[tb]
\centering
\begin{minipage}[c]{0.25\textwidth}
\includegraphics[width=.9\textwidth]{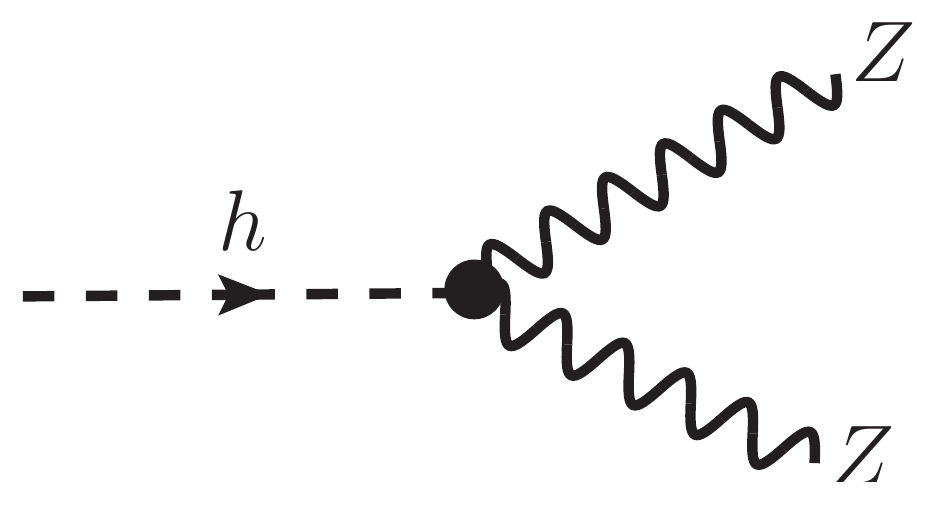}
\end{minipage}
\begin{minipage}[c]{0.3\textwidth}
\includegraphics[width=.9\textwidth]{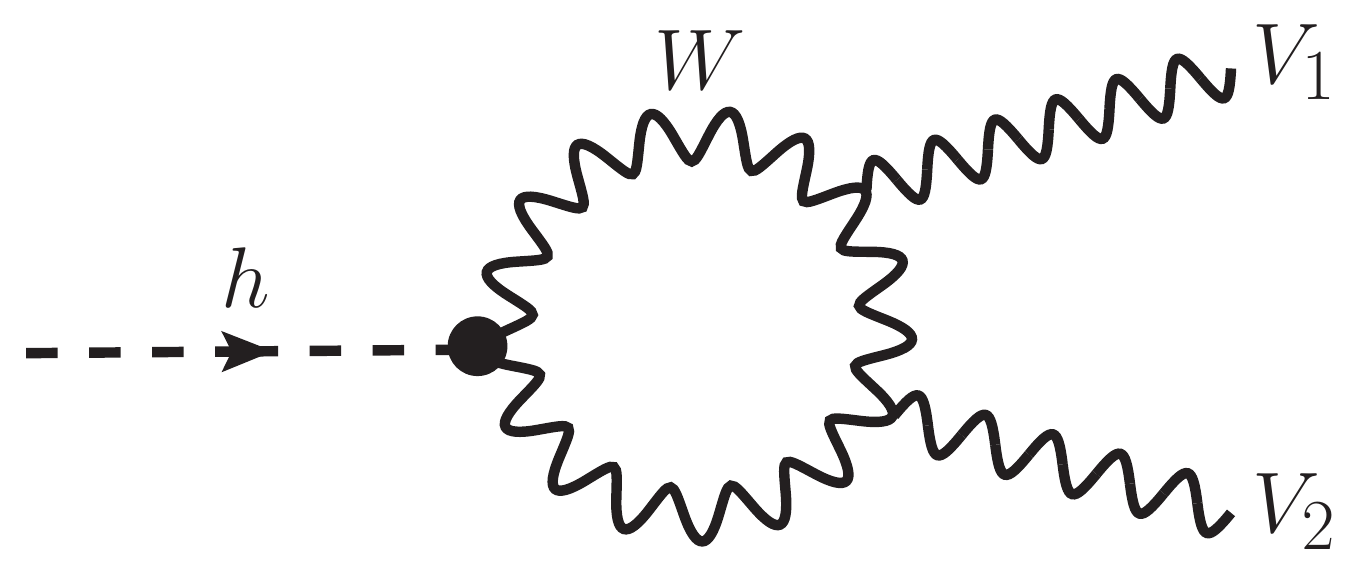}
\end{minipage}
\begin{minipage}[c]{0.3\textwidth}
\includegraphics[width=.9\textwidth]{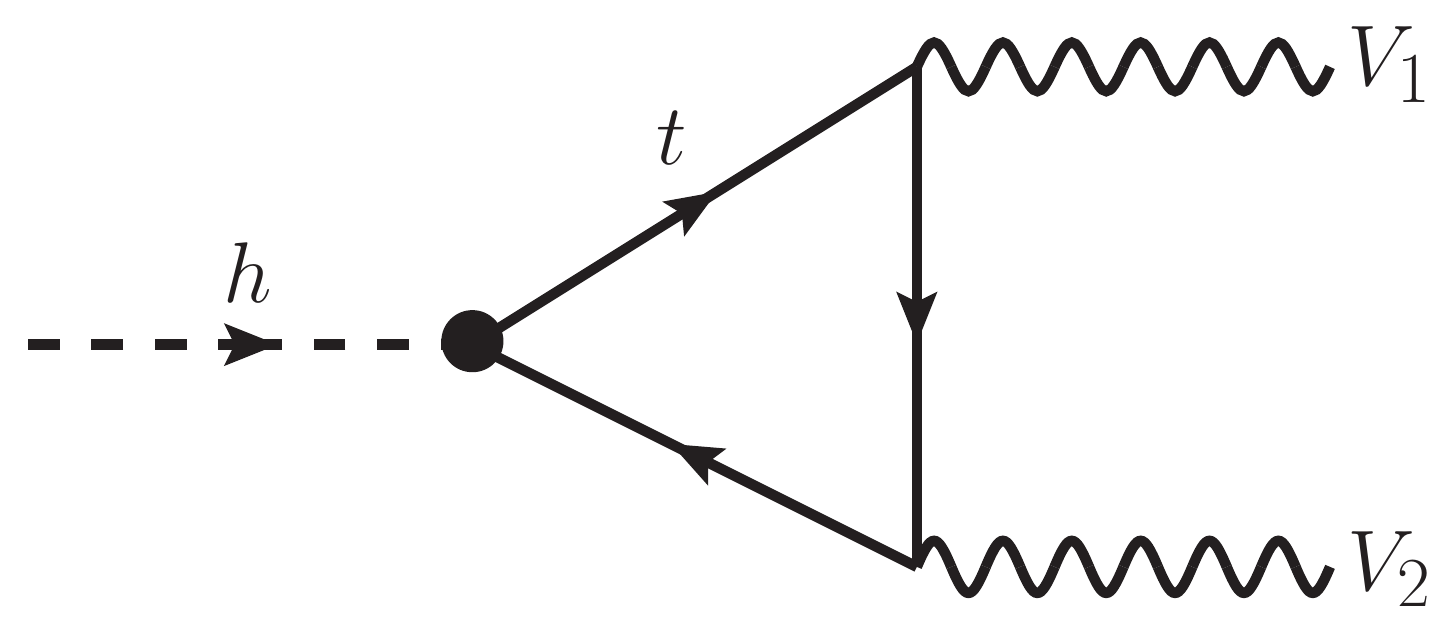}
\end{minipage}
\hfill
\caption{Relevant contributions to the $hVV$ effective couplings mediating Higgs to four lepton decays as shown in~\fref{hto4l}.~On the left is the tree-level contribution mediated by the $Z$ boson pairs, while in the middle and right are example one-loop diagrams mediated by the $W$ boson and top quark, respectively.}
\label{fig:Feyns}
\end{figure*}

The one-loop effects mediated by the top and $W$ loops generate effective couplings to $Z$ and photon pairs as shown schematically in~\fref{hto4l}.~This has been used extensively to study~\cite{Nelson:1986ki,Soni:1993jc,Chang:1993jy,Barger:1993wt,Arens:1994wd,Choi:2002jk,Buszello:2002uu,Godbole:2007cn,Kovalchuk:2008zz,Cao:2009ah,Gao:2010qx,DeRujula:2010ys,Gainer:2011xz,Campbell:2012cz,Campbell:2012ct,Belyaev:2012qa,Coleppa:2012eh,Bolognesi:2012mm,Boughezal:2012tz,Stolarski:2012ps,Avery:2012um,Chen:2012jy,Modak:2013sb,Gainer:2013rxa,Grinstein:2013vsa,Sun:2013yra,Anderson:2013fba,Chen:2013waa,Buchalla:2013mpa,Chen:2013ejz,Gainer:2014hha,Chen:2014gka,Chen:2014pia,Chen:2015iha,Bhattacherjee:2015xra,Gonzalez-Alonso:2015bha} the spin of the Higgs as well as the $CP$ and tensor structure of its coupling to gauge boson pairs.~These effective couplings can be parameterized by the higher dimensional operators:
\bea
\mathcal{L} \supset 
 \frac{h}{4v}
\Big( c_{\gamma} F^{\mu\nu}F_{\mu\nu} + 2c_{Z\gamma} Z^{\mu\nu}F_{\mu\nu} + c_{Z} Z^{\mu\nu}Z_{\mu\nu} 
 \Big),~
\label{eq:eff-op}
\eea
where $F^{\mu\nu}$ ($Z^{\mu\nu}$) is the usual field strength tensor for the photon ($Z$).~If the Higgs has axial couplings to the top quark, CP odd operators with dual field strengths can also be generated.~However, these will have little effect on the sensitivity to $\lambda_{WZ}$ since they are weakly correlated with the CP even operators into which $\lambda_{WZ}$ enters.

The operators in~\eref{eff-op} capture the leading dependance in inverse powers of the loop particle mass, with higher order corrections being quite small~\cite{Gonzalez-Alonso:2014eva} over much of the phase space for a 125~GeV Higgs boson.~The momentum dependence can in principle be relevant in certain regions of phase space and factoring it in may aid in sensitivity, warranting closer examination once higher experimental precision is achieved.~For present purposes in this initial study of probing $\lambda_{WZ}$, it is sufficient for us to consider the form factors as constant in the loop momentum as done in~\cite{Chen:2015rha}.~In this case, and assuming an on-shell Higgs decay as well as fixed loop particle masses, the form factors are then functions only of the loop particle couplings to the Higgs boson, specifically $g_W$ and the top Yukawa coupling.

As has been emphasized in~\cite{Chen:2013ejz,Chen:2014gka,Chen:2014pia,Chen:2015iha,Chen:2015rha}, due to having the same propagator structure as the tree-level amplitude, the sensitivity to the higher dimensional $hZZ$ effective couplings in~\eref{eff-op} is significantly weaker than for the $hZ\gamma$ and in particular $h\gamma\gamma$ effective couplings~\cite{Chen:2015iha}.~Furthermore, though the $hZZ$ effective couplings receive contributions from top and $W$ loops, there are also a number of other one-loop contributions involving $Z$ and Higgs bosons~\cite{Bredenstein:2006rh,Bredenstein:2006nk,Boselli:2015aha}.~The already weak sensitivity to these $hZZ$ effective couplings makes disentangling the contributions involving $g_Z$ and $g_W$ from other loop contributions to these couplings difficult.~We will therefore simply model these loop effects with the $c_{Z}$ effective coupling\footnote{We have also included the additional effective $ZZ$ couplings (including CP odd ones) considered in~\cite{Chen:2015iha}, but as they have only a small effect on our results we do not discuss them explicitly.}~in~\eref{eff-op} and treat it as a nuisance parameter in our statistical analysis of $\lambda_{WZ}$.~As was the case for the top Yukawa~\cite{Chen:2015rha} and will be shown below, once sufficient statistics are obtained, the sensitivity to $\lambda_{WZ}$ is negligibly affected by whether or not we allow these $ZZ$ effective couplings to vary.~Thus, as was also the case for the top Yukawa~\cite{Chen:2015rha}, the sensitivity to $\lambda_{WZ}$ is dominated by the contributions entering via $c_{Z\gamma}$ and especially $c_{\gamma}$ in~\eref{eff-op}.

Unlike the top Yukawa couplings, the tree level coupling $g_{W}$ will also enter into other one-loop diagrams~\cite{Bredenstein:2006rh,Bredenstein:2006nk} involving the $W$ boson that cannot be parameterized by the operators of the type in \eref{eff-op}, such as pentagon diagrams with photons connecting the final-state leptons or amplitudes involving box diagrams.~There are also real photon emission effects in $h\to 4\ell$~\cite{Bredenstein:2006rh,Bredenstein:2006nk,Boselli:2015aha} which can be non-negligible in certain regions of phase space, but which can also be included~\cite{Gonzalez-Alonso:2014eva}.~The key point however is that, in addition to being suppressed over most of the phase space~\cite{Gonzalez-Alonso:2014eva}, these corrections do not contribute at one loop to the $Z\gamma$ or $\gamma\gamma$ effective couplings in~\eref{eff-op} to which we are most sensitive in $h\to4\ell$~\cite{Chen:2014gka,Chen:2015iha}.~We thus neglect them in this preliminary study, but including them may further aid in sensitivity and a detailed investigation of their effects will be worthwhile once enough data is obtained for higher precision measurements.

After the $W$ and top loops, the next largest contribution to the effective $Z\gamma$ and $\gamma\gamma$ couplings comes from the bottom quark contribution.~This effect is suppressed by $\sim (m_b/m_t)^2$ in the matrix element relative to the top contribution, which is itself subdominant to the $W$ loop.~Thus, to a sufficiently good approximation, the $Z\gamma$ and $\gamma\gamma$ effective couplings \emph{only} receive contributions at this order from the $W$ boson and top quark loops in~\fref{Feyns}.~Furthermore, in the limit of negligible loop momentum effects in which we work, these are given simply by the effective couplings which control Higgs decays to \textit{on-shell} $\gamma\gamma$ and $Z\gamma$ pairs.~These one-loop contributions have been computed for $h\to Z\gamma$~\cite{Cahn:1978nz,Bergstrom:1985hp} and $h\to\gamma\gamma$~\cite{Ellis:1975ap,Shifman:1979eb} (including pseudoscalar couplings~\cite{Weiler:1988xn} for the top) and can be straightforwardly incorporated~\cite{Chen:2015rha} into the analytic expressions for the $h\to4\ell$ fully differential cross section computed in~\cite{Chen:2012jy,Chen:2013ejz}.~With this we can go on to perform various statistical tests to assess the possibility of probing $\lambda_{WZ}$ in $h\to4\ell$ decays.~More details of this implementation can be found in~\cite{Chen:2015rha}.

%%%%%%%%%%%%%%%
\subsection{Comments on top Yukawa sector}

As has been discussed, though we are probing $\lambda_{WZ}$ through the loop generated $hZ\gamma$ and $h\gamma\gamma$ effective couplings, there is generically also a top quark contribution.~When attempting to establish $\lambda_{WZ}$, we can assess the sensitivity given specific assumptions about the top Yukawa couplings.~Ideally however, we would like to probe $\lambda_{WZ}$ and, in particular, establish its overall sign independently of the top Yukawa sector.

The top quark couplings to the Higgs can be parametrized generically with the operators,
\bea
\label{eq:Ylag}
\mathcal{L}_t \supset \frac{m_t}{v} h \bar{t} ( y_t  + i \tilde{y}_t \gamma^5 )t ,
\eea
where both CP even ($y_t$) and CP odd couplings ($\tilde{y}_t$) are present and can in principle be positive or negative.~The presence of $\tilde{y}_t$ however, will have little effect on our results for the sensitivity to $\lambda_{WZ}$.~This is because it enters only into the CP odd $hV\gamma$ effective couplings which, as discussed, are weakly correlated with the CP even effective couplings (see~\eref{eff-op}) into which $\lambda_{WZ}$ enters.~Thus, though we include CP odd couplings as nuisance parameters in our statistical analysis, we do not discuss them further here, but see~\cite{Chen:2015rha} for a detailed discussion. 

The CP even Yukawa coupling $y_t$ on the other hand does enter into the same CP even $hVV$ effective couplings in~\eref{eff-op} and is therefore highly correlated with $\lambda_{WZ}$.~Thus, though we are not assessing the sensitivity to the top Yukawa sector~\cite{Chen:2015rha}, how it is treated during our statistical analysis is crucial for assessing the ability to probe $\lambda_{WZ}$ in a model independent way using $h\to4\ell$ decays.~After inputing the numerical values for couplings and masses, the CP even $c_{Z\gamma}$ and $c_{\gamma\gamma}$ effective couplings generated by the top and $W$ loops\footnote{Some sources, including~\cite{Gunion:1989we}, have a sign flip in the definition of the two effective couplings.~Here, we use a natural convention for the couplings based on the effective operators in~\eref{eff-op} and emphasize that with this convention they have the same sign in the SM.~This is in contrast to using the tensor structure to define the phase in which a sign difference between the couplings can arise.~See Appendix B of~\cite{Hartling:2014zca} for a more detailed explanation.} are given by~\cite{Cahn:1978nz,Bergstrom:1985hp,Ellis:1975ap,Shifman:1979eb},
\bea
\label{eq:cVAvalues}
c_\gamma &\simeq& -0.01 \, g_W + 0.002 \, y_t \nn\\
c_{Z\gamma} &\simeq& -0.015 \,g_W + 0.0008 \, y_t .
\eea
Here we see explicitly the way $g_W$ and $y_t$ are correlated and also the fact that having access to both the $c_{\gamma}$ and $c_{Z\gamma}$ effective couplings in $h\to4\ell$ allows us to lift degeneracies in the $(g_W, y_t)$ plane.~Clearly then, the assumptions we make about $y_t$ will greatly affect the sensitivity to $g_W$ and thus the ratio of couplings $\lambda_{WZ}$.~Generically the couplings in~\eref{Ylag} can be treated as free nuisance parameters or fixed under certain assumptions, both of which we consider in this study.

When treating the Yukawa couplings as nuisance parameters, we will make only the relatively conservative assumption that the \emph{magnitude} of the top Yukawa couplings is $\lesssim \mathcal{O}(2)$, but otherwise we treat them as free parameters (allowed to be positive or negative) before being integrated over.~While larger Yukawa couplings could in principle be allowed, this can quickly lead to vacuum stability issues~\cite{EliasMiro:2011aa,Alekhin:2012py,Degrassi:2012ry} whether they are positive or negative~\cite{Reece:2012gi}.~In addition, larger couplings risks running into conflict with experimental constraints~\cite{Khachatryan:2014ira,Khachatryan:2014qaa,Corbett:2015ksa}.

%%%%%%%%%%%%%%%
\subsection{Comments on dominant $4\ell$ background}

We have included the dominant irreducible non-Higgs background coming from the continuum $q\bar{q}\to4\ell$ process~\cite{Khachatryan:2014kca,Aad:2015rka}.~Though it enters almost entirely due to detector resolution~\cite{Chen:2015iha}, it can have non-negligible effects.~Thus in our statistical analysis the background fraction in a given data set is an additional parameter in the likelihood which must be accounted for.~For this $q\bar{q}\to4\ell$ background, we utilize the analytic expressions computed in~\cite{Chen:2012jy,Chen:2013ejz} and follow the procedure in~\cite{Gainer:2011xz,Chen:2015iha,Chen:2015rha} to build the complete signal plus background likelihood for both the $2e2\mu$ and $4e/4\mu$ channels.~To approximate these resolution effects we have implemented a crude modelling of detector resolution\footnote{For a more realistic analysis, careful treatment of detector resolution and additional background effects can be done with the framework in~\cite{Chen:2014pia,Chen:2014hqs,Khachatryan:2014kca}, but is beyond our scope here and left to future work.~However, these detector effects are not expected to qualitatively change our results~\cite{Chen:2014pia}.} as described in~\cite{Chen:2015iha}.

We have also included the relevant phase space selection efficiencies which is necessary to accurately determine the  background fraction.~Since the efficiency depends on the $p_T$ and rapidity of the $4\ell$ system, this necessitates the inclusion of the dominant signal and background production mechanisms.~For this we have utilized a FeynRules/MadGraph5\_aMC@NLO~\cite{Christensen:2008py,Alwall:2014hca} implementation of the effective couplings in~\eref{eff-op} which includes the relevant parton distribution functions (CTEQ6l1~\cite{Lai:1999wy,Pumplin:2002vw}).~Note however that in the final likelihood, production variables are not included and thus only decay information is used.~Since we are interested only in ratios of couplings and not absolute magnitudes, this makes our analysis largely free of many of the uncertainties associated with Higgs production.

In computing the selection efficiencies we have considered both `CMS-like' phase space cuts~\cite{CMS-PAS-HIG-14-014,Khachatryan:2014kca} and a set of `Loose' cuts as defined in~\cite{Chen:2015iha,Chen:2015rha} which greatly relax the requirement on the lepton pair invariant masses, allowing them to be as low as $4$~GeV.\footnote{These cuts were introduced in~\cite{Chen:2015iha,Chen:2015rha}, where they are referred to as `Relaxed - $\Upsilon$' cuts because they cut away the region of phase space in a small window around the $\Upsilon$ resonances at $\sim 9~\rm{GeV}$.}~Though these Loose cuts have been shown to increase the sensitivity in $h\to4\ell$~\cite{Chen:2015iha,Chen:2015rha}, for many of the results presented here the conclusions are not qualitatively affected by which cuts are used so we do not discuss them in detail further.~More details on the implementation of these cuts as well as the construction of the likelihood can be found in~\cite{Gainer:2011xz,Chen:2012jy,Chen:2013ejz,Chen:2015iha} and in particular~\cite{Chen:2015rha} whose procedure we follow closely here when building the signal plus background likelihood.

%%%%%%%%%%%%%%%
%%%%%%%%%%%%%%%
%%%%%%%%%%%%%%%

%%%%%%%%%%%%%%%%%%%%%%%%%%%%%
\section{Probing Electroweak Symmetry Breaking
at the LHC}
\label{sec:lhc}

We now examine different techniques to probe $\lambda_{WZ}$ with $h\to 4\ell$ decays at a 13~TeV LHC.~We first perform a likelihood shape analysis of $h\to4\ell$ decays to estimate how much data will be needed to establish the overall sign of $\lambda_{WZ}$.~We then perform hypothesis testing to assess the ability to distinguish between the two predictions of custodial symmetry for $\lambda_{WZ}$ shown in~\eref{lam}.~Finally, we perform parameter extraction in order to assess more generally the sensitivity to $\lambda_{WZ}$ and estimate how much data will be needed to begin making precision measurements.~More details on the discussion and statistical analysis procedures presented here can be found in~\cite{Gainer:2011xz,Stolarski:2012ps,Chen:2012jy,Chen:2013ejz,Chen:2014gka,Chen:2015iha,Chen:2015rha} to which we refer the interested reader.

%%%%%%%%%%%%%%%
\subsection{Pinning down the sign of $\lambda_{WZ}$}

Our first exploration begins with assessing how much data will be needed to pin down the overall sign of $\lambda_{WZ}$.~As discussed, rate measurements can not determine this sign and furthermore, under the assumption of custodial symmetry it effectively establishes the custodial representation of the Higgs boson.

Following the procedure described in~\cite{Chen:2015rha} we construct the likelihood from the (normalized) signal and background fully differential cross sections.~This likelihood is a function of the couplings $(g_Z, g_W)$ and the set of nuisance parameters $(c_Z, y_t, \tilde{y}_t)$.~A full likelihood is built for each pseudodataset, and integrated\footnote{During this integration, we include a flat prior probability distribution function for $y_t$ which restricts its range to the (perturbative) values $-2 \lesssim y_t \lesssim 2$, but where the boundaries are made `soft' by placing gaussian tails centered at the two endpoints with $\sigma = 0.5$.~More details on this procedure will be given in~\cite{followup3}.} over the different nuisance parameters to obtain the posterior likelihood as a function only of $\lambda_{WZ}$.~When generating pseudodatasets we consider two possibilities.~The first is that the SM is the true underlying model which predicts $\lambda_{WZ} = 1$.~As a second case we also consider the other allowed possibility by custodial symmetry of $\lambda_{WZ} = -1/2$.~An example of the posterior likelihood is shown in~\fref{lamwzLH} for one pseudodataset containing $\mathcal{O}(2000)$ signal events assuming the true underlying model is the SM.
%%%%%
%%%%%
%%%%%
\begin{figure}
\begin{center}
\includegraphics[width=.45\textwidth]{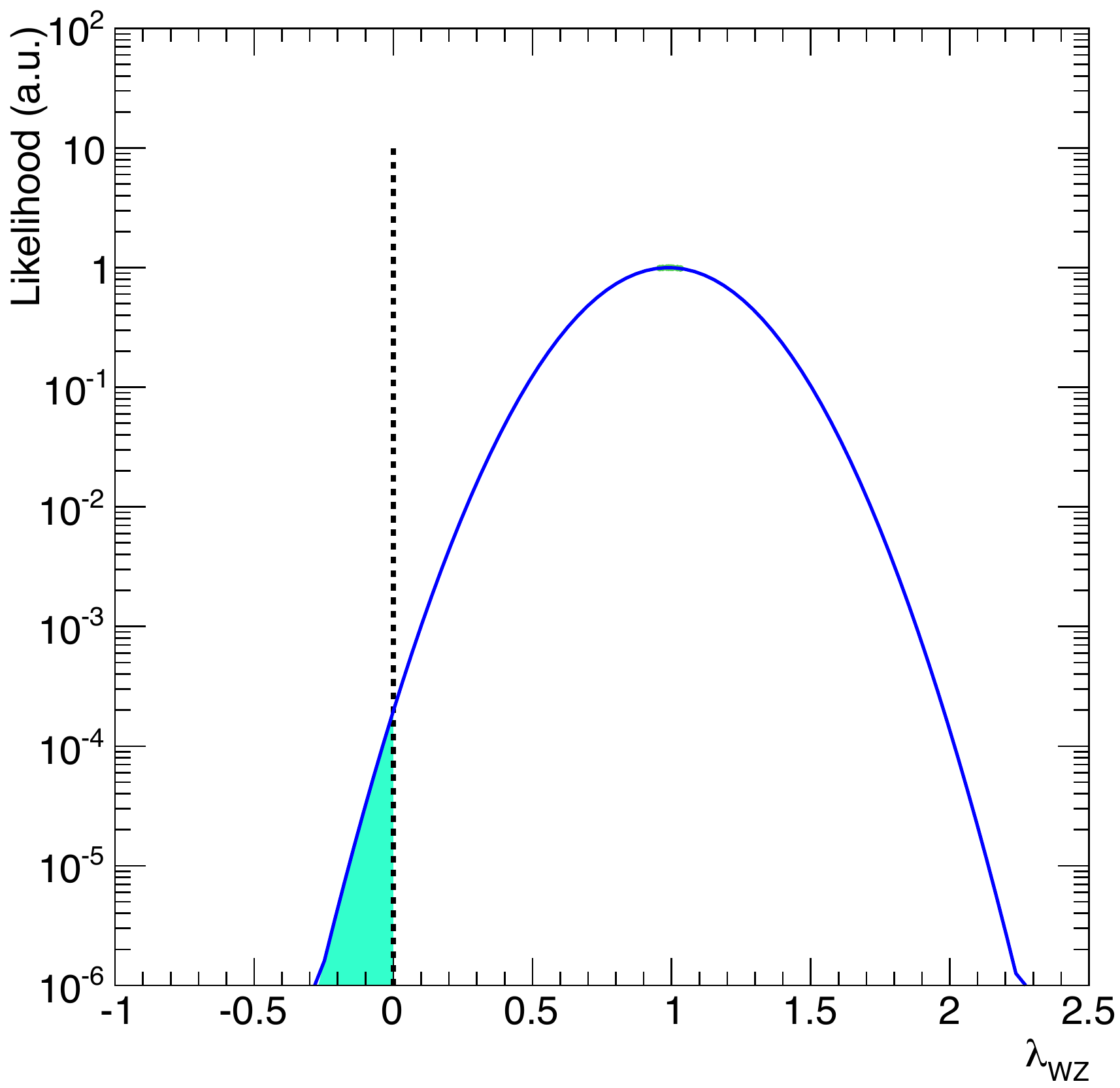}
\end{center}
\caption{Example of the posterior likelihood for one pseudodataset containing $\mathcal{O}(2000)$ signal events and generated for the SM case with $\lambda_{WZ} = 1$.~The shaded turqoise region indicates the area under the curve on the negative side of zero which is translated into a probability (see~\fref{signplot}) that the sign of $\lambda_{WZ}$ is negative.~See text for more details.}
\label{fig:lamwzLH}
\end{figure}
%%%%%
%%%%%
%%%%%

We treat the normalized posterior likelihood as a probability density of the extracted true value of $\lambda_{WZ}$.~Given the observed pseudodataset, we obtain from the posterior likelihood a $p$-value that the true value of $\lambda_{WZ}$ is negative by taking the ratio of the area on the negative side of zero (shaded in turquoise in~\fref{lamwzLH}) to the total area.~For each $p$-value we define a corresponding `effective' $\sigma$ by how much of the tail we have to integrate a (normalized) gaussian to get an equivalent area of $p$ and converting that distance into an effective $\sigma$.~This procedure is then repeated over many pseudodatasets giving a distribution of effective $\sigma$'s which represent the probability that the sign of $\lambda_{WZ}$ is negative.
%%%%%
%%%%%
%%%%%
\begin{figure}
\begin{center}
\includegraphics[width=.44\textwidth]{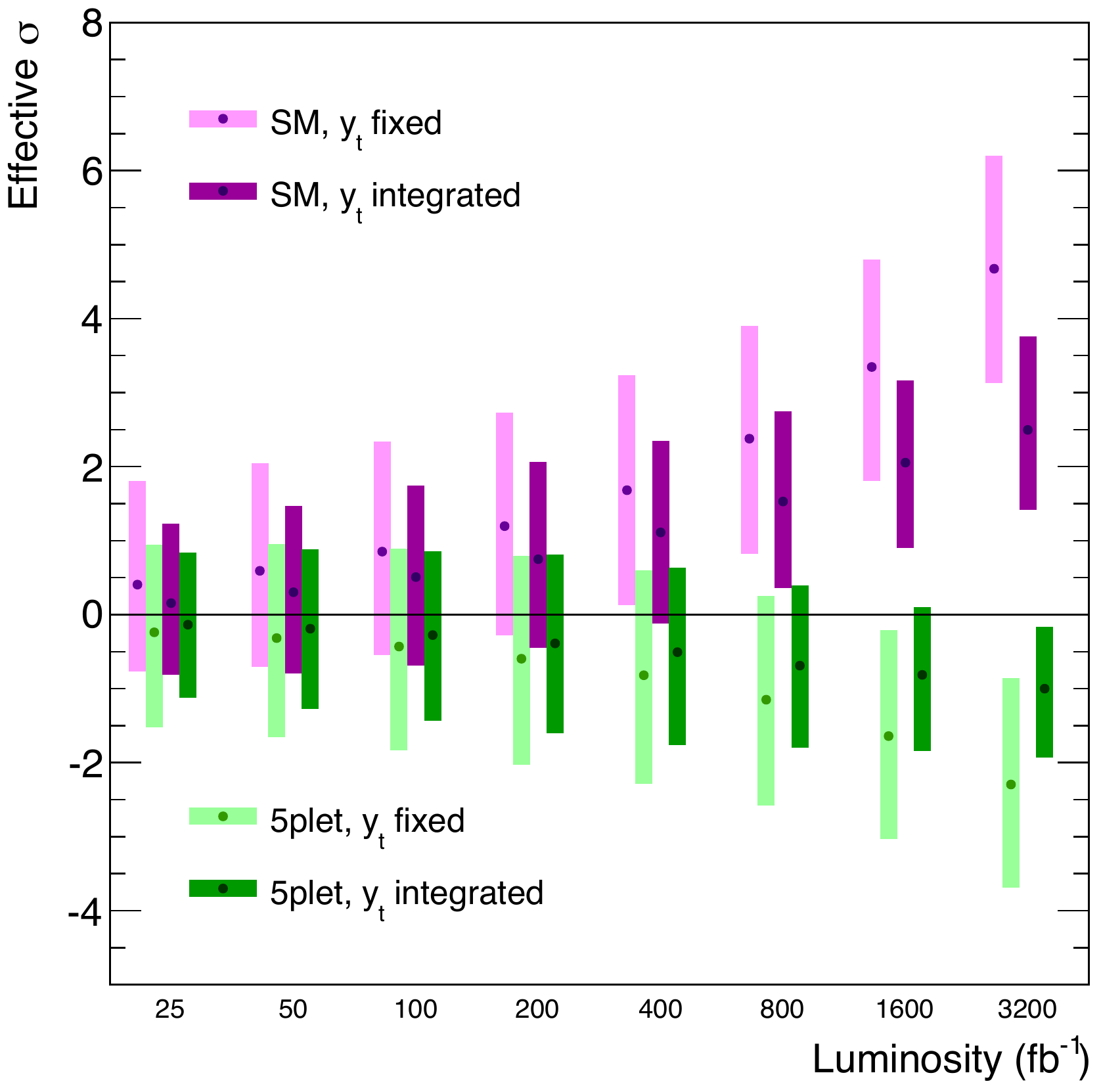}
\end{center}
\caption{Probability, in units of effective $\sigma$'s (see text) as a function of luminosity.~In purple we show the SM, while in green we show the case of a custodial fiveplet.~In the light shaded bands the top Yukawa couplings are fixed to their true values.~In the dark shaded bands the top Yukawa coupling is treated as a nuisance parameter and integrated over assuming a prior probability distribution which restricts its range to perturbative values as discussed in text.~In both cases we have assumed SM production ($gg\to h$ plus VBF at 13 TeV~\cite{Dittmaier:2011ti,Heinemeyer:2013tqa}) times branching ratio and accounted for phase space selection efficiencies as well as the dominant $q\bar{q}\to4\ell$ background.}
\label{fig:signplot}
\end{figure}
%%%%%
%%%%%
%%%%%

In~\fref{signplot} we show the distribution of effective $\sigma$'s for two separate cases.~A negative value for the effective $\sigma$ indicates that the peak of the likelihood is on the negative side of $\lambda_{WZ}$ (as is the case for a custodial fiveplet) whereas a positive value represents a distribution centered on the positive side (as for a custodial singlet).~The dot in each case indicates the median value, and the colored bands represent the central $68.3\%$ interval of the distribution of effective $\sigma$'s.~In purple we show the SM while in green we show the case of a custodial fiveplet.~For both cases we have considered two possibilities.~The first (light shaded bands) is where the top Yukawa couplings are fixed to their true values as predicted in the SM ($y_t = 1$) and for a custodial fiveplet ($y_t = 0$).~In the second case (dark shaded bands) we assume only that the CP even top Yukawa coupling is restricted to perturbative values which, as discussed above, is accomplished via the use of priors.~In order to interpret in terms of luminosity, we have accounted for all relevant selection efficiencies~\cite{CMS-PAS-HIG-15-004,Chen:2015iha} while assuming SM production ($gg\to h$ plus VBF at 13 TeV~\cite{Dittmaier:2011ti,Heinemeyer:2013tqa}) times branching ratio into $4\ell$.

We see in~\fref{signplot} that if the true underlying model is the SM (or something close to it) a 13~TeV LHC should begin directly pinning down the sign of $\lambda_{WZ}$ with as few as $\mathcal{O}(800)~\rm{fb}^{-1}$ of data and corresponding to $\mathcal{O}(1300)$ signal events for the more inclusive Loose cuts.~We also see that, while the sensitivity is degraded, whether we treat the top Yukawa sector as a nuisance parameter which we integrate over or fix it to its true value, it does not qualitatively change this conclusion.

Of course when fixing the top Yukawa to its true value, we can establish the sign of $\lambda_{WZ}$ with more confidence.~We see this in~\fref{signplot} where for the SM case an expected $\sigma \gtrsim 4$ can be obtained at very high LHC luminosities when the top Yukawa is fixed, while only $\sigma \gtrsim 2$ is expected when it is integrated over.~As direct constraints on the top Yukawa from other measurements, such as $tth$ production~\cite{Khachatryan:2014ira,Khachatryan:2014qaa,Corbett:2015ksa} increase in precision and further restrict the upper bound, stronger priors based on experimental data can be used during integration leading to an increase in sensitivity to $\lambda_{WZ}$.~We also see a stronger sensitivity to the SM case than to the custodial fiveplet case.~This is of course because of the smaller value of $|\lambda_{WZ}|$ in the fiveplet case and implies $\mathcal{O}(1)$ or better precision measurements of $\lambda_{WZ}$ can be made in $h\to4\ell$ at the LHC.~We explore this more below.

We thus see that a high luminosity LHC has excellent prospects of directly establishing the overall sign of $\lambda_{WZ}$ in $h\to4\ell$ decays.~As discussed, under the assumption of custodial symmetry this also effectively establishes the custodial nature of the Higgs boson.~Crucially, this does not rely on other measurements of Higgs couplings and is largely independent of assumptions about the top Yukawa sector.

%%%%%%%%%%%%%%
%%%%%%%%%%%%%%
%%%%%%%%%%%%%%

%%%%%%%%%%%%%%%%%%%%%%%%%
\subsection{Testing custodial symmetry}

Motivated by considerations of custodial symmetry in the introduction and the discussion in~\cite{Low:2010jp}, we next consider hypothesis testing to assess how much data is needed to discriminate between the two possibilities for $\lambda_{WZ}$ predicted by custodial symmetry.~As discussed above, this predicts $\lambda_{WZ} = +1$ and $\lambda_{WZ} = -1/2$ for a custodial singlet (as in the SM model) and fiveplet (as in the GM-type models) respectively.~Since custodial symmetry restricts the possibilities to two discrete choices, this makes it particularly suited to hypothesis testing.

We use the hypothesis testing techniques developed in~\cite{DeRujula:2010ys} and utilized in~\cite{Stolarski:2012ps,Falkowski:2014ffa,delaPuente:2016pyh} to construct a test statistic that  measures the separation power between the two models allowed by custodial symmetry.~This is done by constructing the likelihood ratio between the singlet and fiveplet hypotheses.~Pseudoexperiments are then conducted to obtain a distribution of these likelihood ratios.~This is first done assuming the SM as the `true' hypothesis and then repeated assuming the custodial fiveplet is true.~In each case, a distribution of likelihood ratios is obtained after conducting a large set of psuedoexperiments.~The overlap (or lack thereof) between these two distributions can then be converted into a measure of the ability to discriminate between the two models.~We follow closely the procedure in~\cite{Stolarski:2012ps} apart from the additional step that we now include the presence of nuisance parameters which will be integrated over~\cite{followup3} before constructing the likelihood ratio.~Furthermore, here we will present our results in terms of $p$-values instead of $\sigma$'s.

In~\fref{hyptest} we show the probability of mistaking a SM Higgs (custodial singlet) for a custodial fiveplet or \emph{vice versa} as a function of data.~Since there are small (sub percent) differences in selection efficiencies between the singlet and fiveplet, we plot the approximate number of signal events\footnote{Note we are assuming the singlet and fiveplet cases somehow give the same yields prior to analysis selection.~This is akin to asking how much discriminating power is contained \emph{only} in the kinematics of the $4\ell$ events while not using rate information which we assume can be obtained from total cross section measurements.} ($N_S$ on bottom axis) to go along with the luminosity (top axis).~For this we have again assumed SM production (gluon fusion plus VBF) times branching ratio.~For these curves we utilize the `CMS-like' phase space cuts (red) and Loose phase space cuts (blue and green) discussed above and defined in~\cite{Chen:2015rha}.~In the green curve $\lambda_{WZ}$ and the top Yukawa couplings are fixed to their true values.~In the red and blue curves the top Yukawa is again treated as a nuisance parameter and integrated over while restricting its magnitude to be $\lesssim 2$ and fixing $\lambda_{WZ}$ to its true value.
%%%%%
%%%%%
%%%%%
\begin{figure}
\begin{center}
\includegraphics[width=.45\textwidth]{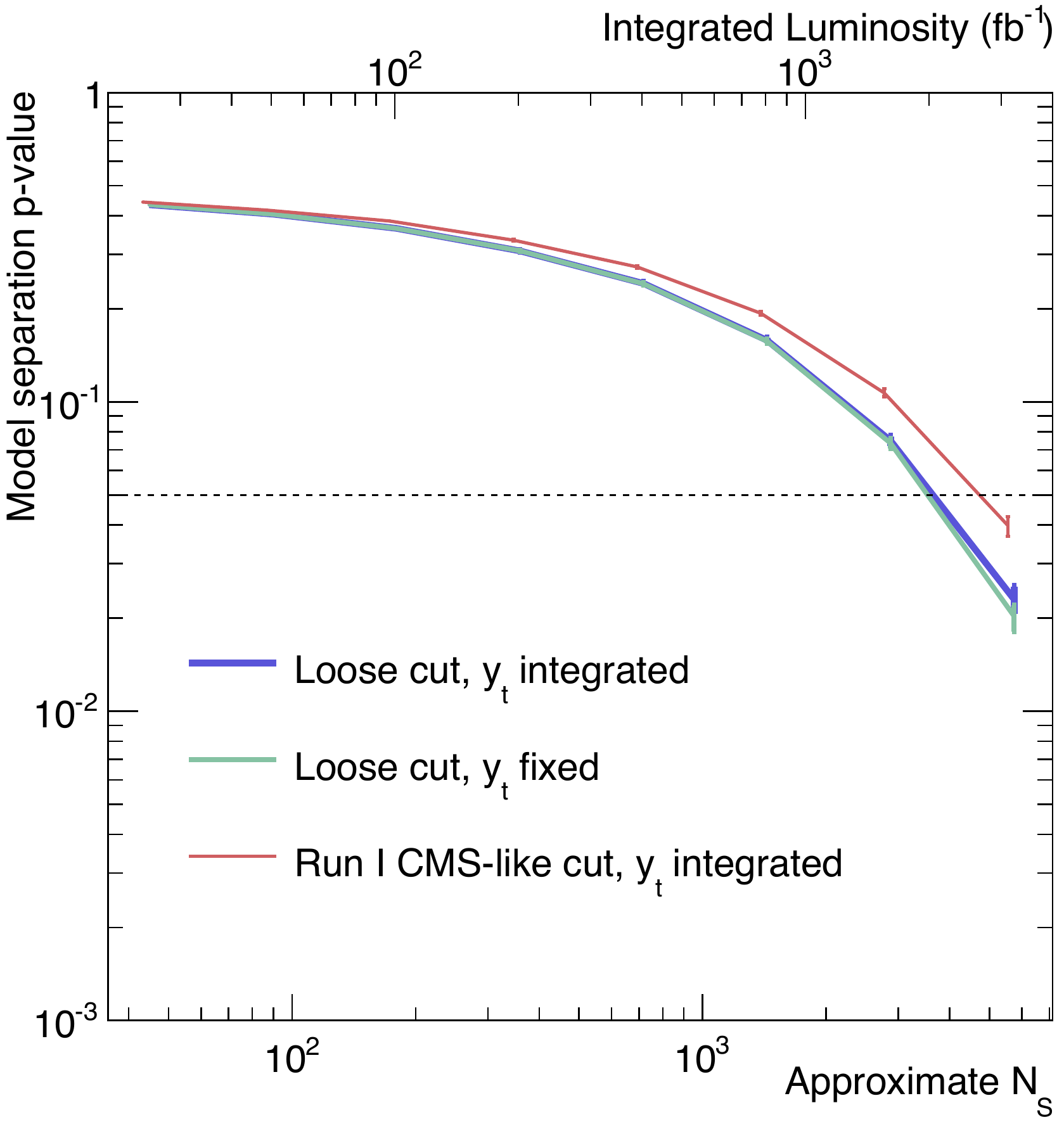}
\end{center}
\caption{Probability of mistaking a Standard Model Higgs (custodial singlet) for a custodial fiveplet or \emph{vice versa}.~For these curves we utilize `CMS-like' phase space cuts (red) and `Loose' phase space cuts (blue and green) as discussed in text.~In the green curve the top Yukawa couplings are fixed to their true values.~In the red and blue curves the top Yukawa coupling is again treated as a nuisance parameter and integrated over as in~\fref{signplot}.~Again we have assumed SM production ($gg\to h$ plus VBF at 13 TeV~\cite{Dittmaier:2011ti,Heinemeyer:2013tqa}) times branching ratio.~Since there are small (sub percent) differences in selection efficiencies between the singlet and fiveplet, we plot an `approximate' $N_S$ along with luminosity.}
\label{fig:hyptest}
\end{figure}
%%%%%
%%%%%
%%%%%

We see that $h\to4\ell$ decays should have the ability to discriminate between the two predictions of custodial symmetry for $\lambda_{WZ}$ at $95\%$ confidence with $\mathcal{O}(3000)$ signal events corresponding to $\gtrsim 2000~\rm{fb}^{-1}$ of data.~We see that the separation power between the two models is not greatly affected by whether we fix the top Yukawa couplings to their true values or treat them as freely varying nuisance parameters.~The red curve also shows the fact that CMS-like cuts are somewhat less sensitive than the more inclusive Loose cuts~\cite{Chen:2015iha} discussed above.

%%%%%%%%%%%%
\subsection{Sensitivity to $|\lambda_{WZ}|$ at LHC and beyond}

Using the parameter extraction methods developed in~\cite{Chen:2013ejz,Chen:2014gka,Chen:2015iha,Chen:2015rha} we examine more generally the sensitivity to $|\lambda_{WZ}|$ as a function of the amount of data.~For this analysis we follow very closely the procedure based on a maximization of the likelihood which is described in~\cite{Chen:2015rha} to which we refer the reader for more details.

To estimate the expected precision we use as our test statistic the \emph{average error} defined in~\cite{Chen:2013ejz,Chen:2014pia,Chen:2015iha} as,
\bea
\label{eq:sigma}
\sigma(\lambda_{WZ}) = \sqrt{\frac{\pi}{2}} \langle |\hat{\lambda}_{WZ} - \bar{\lambda}_{WZ}| \rangle ,
\eea
where $\hat{\lambda}_{WZ}$ is the value of the best fit parameter point obtained by maximization of the likelihood with respect to~$\lambda_{WZ}$.~Here $\bar{\lambda}_{WZ}$ represents the `true' value with which our data sets are generated utilizing a MadGraph5\_aMC@NLO~\cite{Alwall:2014hca} implementation of the effective $hVV$ couplings~\cite{Chen:2012jy,Chen:2013ejz}.~The average error is then found by conducting a large number of pseudoexperiments for a given number of (expected) events\footnote{Each dataset varies in size according to a poisson distribution with mean at the expected number of events for each of the four components ($2e2\mu$ and $4e/4\mu$ for signal and background).} and obtaining a distribution for $\hat{\lambda}_{WZ}$.~This distribution will be centered around some average value with a width that is then translated into our average error as in~\eref{sigma}.~This definition converges to the usual gaussian interpretation of a $\sigma$ when the distribution of $\hat{\lambda}_{WZ}$ is perfectly gaussian.~We repeat this procedure for a range of number of signal events to obtain $\sigma(\lambda_{WZ})$ as a function of $N_S$.

We show in~\fref{curves} sensitivity curves for $\sigma(\lambda_{WZ})$ as a function of the number of signal events (bottom axis) and luminosity $\times$ efficiency (top axis) assuming SM production ($gg\to h$ plus VBF at 13 TeV~\cite{Dittmaier:2011ti,Heinemeyer:2013tqa}).~We fit to a `true' point of $\lambda_{WZ} = 1$ corresponding to the SM prediction and again consider both CMS-like phase space cuts (purple) and Loose phase space cuts (orange, pink, and green) discussed above and defined in~\cite{Chen:2015rha}.~We also show a second luminosity axis assuming the nominal efficiency ($\sim 30\%$) for the Loose cuts.~We again compare the case where the top Yukawa coupling is treated as a nuisance parameter to when it is fixed (blue curve) to its true value.~However, now we consider two separate cases when treating it as a nuisance parameter.

%%%%%
%%%%%
%%%%%
\begin{figure}
\begin{center}
\includegraphics[width=.45\textwidth]{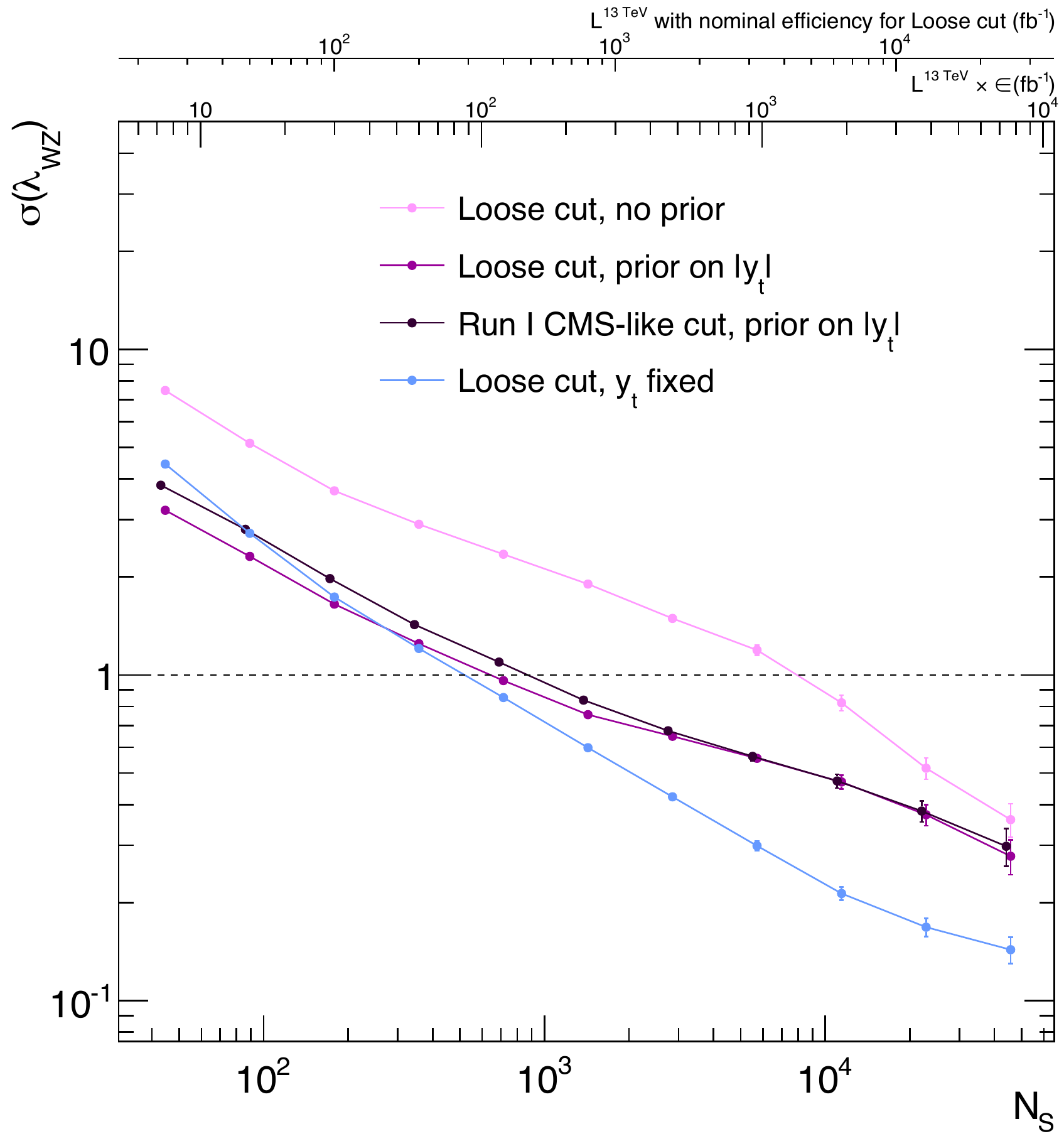}
\end{center}
\caption{Sensitivity curves for the `average error' $\sigma(\lambda_{WZ})$ defined in~\eref{sigma} versus the number of signal events (bottom axis) and luminosity $\times$ efficiency (top axis) for which we assume SM production ($gg\to h$ plus VBF at 13 TeV~\cite{Dittmaier:2011ti,Heinemeyer:2013tqa}) times branching ratio.~We also show a second luminosity axis assuming the nominal efficiency ($\sim 30\%$) for the Loose cuts discussed in text.~In all cases we fit to a `true' point of $\lambda_{WZ} = 1$ as found in the SM and include the dominant $q\bar{q} \to 4\ell$ background.~See text for more information.}
\label{fig:curves}
\end{figure}
%%%%%
%%%%%
%%%%%

In the first case (pink and purple), we again use a prior to restrict the magnitude of the top Yukawa coupling to the physically viable values of $y_t\lesssim 2$ but otherwise allow it to vary freely.~In the second case, no prior is applied to the top Yukawa sector and the couplings are allowed to take on any value when varied during the likelihood maximization procedure, including very large positive or negative values.~Since $y_t$ is highly correlated with $\lambda_{WZ}$, this introduces large errors at small data sets.~Though we do not consider this case as physically viable, we show it for comparison and in particular, to demonstrate that eventually the sensitivity to $\lambda_{WZ}$ does not depend on whether a prior on $y_t$ is used.~We see this at very high luminosity where the curves with and without priors begin to converge.

We see in~\fref{curves} that $\mathcal{O}(1)$ precision on $\lambda_{WZ}$ may be achievable with $\sim 500-800$ signal events depending on whether the top Yukawa is fixed to its true value or treated as a nuisance parameter with a prior restricting it to physically viable values.~Assuming $100\%$ efficiency and SM production rates, this would require $\lesssim 100~\rm{fb}^{-1}$, though more realistically $\sim 300~\rm{fb}^{-1}$ will be needed.~We also see as expected that once sufficient statistics are achieved, the sensitivity is much stronger when the top Yukawa couplings are fixed.

This also implies that as direct measurements of the top Yukawa coupling become more precise, such as from $tth$ production~\cite{Khachatryan:2014ira,Khachatryan:2014qaa}, more restrictive priors on the top Yukawa can be used to enhance the sensitivity to $\lambda_{WZ}$ close to that achievable when fixing $y_t$ (blue curve).~In this case, the ultimate LHC sensitivity could reach $\mathcal{O}(20\% - 30\%)$ again assuming $100\%$ selection efficiencies and $\sim 3000~\rm{fb}^{-1}$.~While the LHC may be able to achieve moderately high precision, clearly measurements of $\lambda_{WZ}$ with $h\to4\ell$ decays would benefit greatly from a future 100~TeV collider where gluon fusion production would be more than an order of magnitude larger~\cite{Arkani-Hamed:2015vfh}.

%%%%%%%%%%%%%%%%%%%%%%%%%%%%%
\section{Conclusions}
\label{sec:conclusion}

We have examined the possibility of using Higgs to four lepton decays to study the ratio of its couplings to $WW$ and $ZZ$ pairs as defined in~\eref{hLag} and~\eref{LWZ}.~This ratio, $\lambda_{WZ}$, is a crucial parameter of electroweak symmetry breaking and a probe of the well known `custodial' symmetry of the gauge boson mass matrix.~Utilizing all observables in the normalized fully differential decay width, we have constructed a matrix element analysis in order to perform various statistical tests including:~likelihood shape analysis, hypothesis testing, and parameter extraction to assess the ability of $h\to4\ell$ decays to probe the magnitude and phase of $\lambda_{WZ}$.

In particular, we have emphasized that this channel is a uniquely effective probe of the sign of $\lambda_{WZ}$.~Furthermore, we have emphasized that simply establishing the overall sign of $\lambda_{WZ}$ effectively determines the custodial representation of the Higgs boson.~We have performed a likelihood shape analysis to estimate how much data will be needed to establish the sign at a 13~TeV LHC.~We find that $h\to4\ell$ decays will begin to become sensitive to the overall sign of $\lambda_{WZ}$ with as few as $\mathcal{O}(800)~\rm{fb}^{-1}$ assuming SM production and decay rates.~As additional data is collected the LHC should be able to firmly establish the overall sign, and therefore the custodial nature of the Higgs boson, well before the end of running at a high luminosity phase.

We have also performed hypothesis testing to assess the ability to discriminate between the two predictions of custodial symmetry for $\lambda_{WZ}$ in~\eref{lam} and find they can be distinguished at $95\%$ confidence with $\gtrsim 2000~\rm{fb}^{-1}$ of data, again assuming SM production and decay rates.~In addition, we have performed parameter extraction in order to assess the sensitivity to $\lambda_{WZ}$ and estimate how much data will be needed to begin making precision measurements.~We find that a precision of $\mathcal{O}(20 - 50)\%$ can potentially be achieved by end of high luminosity LHC running.~We have also shown how the precision can be improved by utilizing prior probability distribution functions based on experimental and theoretical constraints on the top Yukawa couplings. 

Finally, we have emphasized the ability of $h\to4\ell$ decays to probe $\lambda_{WZ}$ without relying on other measurements of the Higgs boson couplings to electroweak vector bosons or the top quark.~This makes $h\to4\ell$ decays a unique and independent probe of electroweak symmetry breaking at the LHC and beyond at future colliders.

%%%%%%%%%%%%%%%%%%%%%%%%%%%%%%%%%%%%%%
~\\
\noindent
{\bf Acknowledgments:}~We would like to thank Francisco del Aguila,~Adam Falkowski,~Heather Logan,~Ian Low,~Jose Santiago, and~Martin Schmaltz for useful discussions.~R.V.M.~also thanks Fermilab and the Mainz Institute for Theoretical Physics (MITP) for their hospitality and partial support during the completion of this work as well as Tilman Plehn and the participants of the FIRSTRESULTS2016 workshop for useful discussions.~The work of Y.C.~is supported by the Weston Havens Foundation and DOE grant No.~DE-FG02-92-ER-40701.~D.S.~is supported in part by the Natural Sciences and Engineering Research Council of Canada (NSERC).~R.V.M.~is supported by MINECO, under grant number FPA2013-47836-C3-2-P.
%\clearpage 
 
\bibliographystyle{apsrev}
\bibliography{references}

\end{document}